\documentclass[journal]{IEEEtran}

\usepackage[T1]{fontenc}
\usepackage{graphicx}
\usepackage{cite}
\usepackage{subcaption}

\usepackage{lipsum}
\usepackage{overpic}
\usepackage{amssymb}
\usepackage{amsmath}
\usepackage{amsthm}
\usepackage{array}
\usepackage{color}
\usepackage{url}

\IEEEoverridecommandlockouts

\theoremstyle{plain}

\newtheorem{definition}{Definition}
\newtheorem{lemma}{Lemma}

\newtheorem{corollary}{Corollary}
\newtheorem{remark}{Remark}
\newtheorem{proposition}{Proposition}

\newcommand{\vect}[1]{\mathbf{#1}}

\def\diag{\mathrm{diag}}

\def\kron{\otimes}
\def\tr{\mathrm{tr}}
\def\rank{\mathrm{rank}}
\def\Htran{\mbox{\tiny $\mathrm{H}$}}
\def\Ttran{\mbox{\tiny $\mathrm{T}$}}
\def\CN{\mathcal{N}_{\mathbb{C}}} 
\def\imagunit{\mathsf{j}} 

\def\sinc{\mathrm{sinc}}
\def\dV{d_{\mathrm{V}}}
\def\dH{d_{\mathrm{H}}}

\begin{document}

\title{\huge Rayleigh Fading Modeling and Channel Hardening for Reconfigurable Intelligent Surfaces}

\author{\IEEEauthorblockN{Emil Bj{\"o}rnson, \emph{Senior Member, IEEE}, Luca Sanguinetti, \emph{Senior Member, IEEE}\vspace{-0.5cm}
\thanks{\copyright 2020 IEEE. Personal use of this material is permitted. Permission from IEEE must be obtained for all other uses, in any current or future media, including reprinting/republishing this material for advertising or promotional purposes, creating new collective works, for resale or redistribution to servers or lists, or reuse of any copyrighted component of this work in other works.\newline
\indent E. Bj{\"o}rnson was supported by ELLIIT and the Wallenberg AI, Autonomous Systems and Software Program (WASP). L. Sanguinetti was supported by the University of Pisa under the PRA 2018-2019 Research Project CONCEPT, and by the Italian Ministry of Education and Research (MIUR) in the framework of the CrossLab project (Departments of Excellence).\newline
\indent E.~Bj\"ornson is with the Department of Computer Science, KTH Royal Institute of Technology, 10044 Stockholm, Sweden, and Link\"{o}ping University, 58183 Link\"{o}ping, Sweden (emilbjo@kth.se). L.~Sanguinetti is with the University of Pisa, Dipartimento di Ingegneria dell'Informazione, 56122 Pisa, Italy (luca.sanguinetti@unipi.it).}
}}
\maketitle

\begin{abstract}
A realistic performance assessment of any wireless technology requires the use of a channel model that reflects its main characteristics. The independent and identically distributed Rayleigh fading channel model has been (and still is) the basis of most theoretical research on multiple antenna technologies in scattering environments. This letter shows that such a model is not physically appearing when using a reconfigurable intelligent surface (RIS) with rectangular geometry and provides an alternative physically feasible Rayleigh fading model that can be used as a baseline when evaluating RIS-aided communications. The model is used to revisit the basic RIS properties, e.g., the rank of spatial correlation matrices and channel hardening.
\end{abstract}


\vspace{-2mm}

\IEEEpeerreviewmaketitle

\section{Introduction}

Reconfigurable intelligent surface (RIS) is an umbrella term used for two-dimensional surfaces that can reconfigure how they interact with electromagnetic waves \cite{Renzo2020b}, to synthesize the scattering and absorption properties of other objects. This feature can be utilized to improve the wireless physical-layer channel between transmitters and receivers; for example, to enhance the received signal power at desired locations and suppress interference at undesired locations \cite{Wu2019a}.
The RIS technology can potentially be implemented using software-defined metasurfaces \cite{Tsilipakos2020a}, which consist of many controllable sub-wavelength-sized elements. The small size makes each element act as an almost isotropic scatterer and the RIS assigns a pattern of phase-delays to the elements to create constructive and destructive interference in the desired manner \cite{Ozdogan2019a}.

When analyzing new physical-layer technologies, it is a common practice to consider the tractable independent and identically distributed (i.i.d.) Rayleigh fading channel model. For example, the basic features of Massive MIMO (multiple-input multiple-output) were first established using that model \cite{Marzetta2016a} and later extended to spatially correlated channels \cite{massivemimobook}. 
Only physically feasible channel models can provide accurate insights, but the i.i.d.~Rayleigh fading model can be observed in practice if a half-wavelength-spaced uniform linear array (ULA) is deployed in an isotropic scattering environment \cite{Marzetta2016a}.

Several recent works have analyzed RIS-aided communications under the assumption of i.i.d.~Rayleigh fading \cite{Wu2019a,Yu2019a,Huang2018a}. In this letter, we prove that this fading distribution is not physically appearing when using an RIS in an isotropic scattering environment, thus it should not be used. Motivated by this observation, we derive a spatially correlated Rayleigh fading model that is valid under isotropic scattering. We analyze the basic properties of the new model, including how the rank of the correlation matrices depends on the physical geometry. We also define a new channel hardening concept and prove when it is satisfied in RIS-aided communications.

\textit{Reproducible Research:} The simulation code is available at: \url{https://github.com/emilbjornson/RIS-fading}

\section{System Model}

We consider a single-antenna transmitter communicating with a single-antenna receiver in an isotropic scattering environment, while being aided by an RIS equipped with $N$ reconfigurable elements. The received signal $r \in \mathbb{C}$ is \cite{Huang2018a}
\begin{equation} \label{eq:received-signal1}
	r =  \left( \vect{h}_2^{\Ttran} \boldsymbol{\Phi} \vect{h}_1 + h_{\mathrm{d}} \right) s + w
\end{equation}
where $s$ is the transmitted signal with power $P=\mathbb{E}\{ |s|^2\}$ and $w \sim \CN(0,\sigma^2)$ is the noise variance. The configuration of the RIS is determined by the diagonal phase-shift matrix $\boldsymbol{\Phi} = \diag(e^{-\imagunit \phi_1},\ldots,e^{-\imagunit \phi_N})$. The direct path $h_{\mathrm{d}} \in \mathbb{C}$ has a Rayleigh fading distribution due to the isotropic scattering assumption \cite{Marzetta2016a}: $h_{\mathrm{d}} \sim \CN(0,\beta_d)$ where $\beta_d$ is the variance.

A main goal of this paper is to characterize the fading distribution of the channel $\vect{h}_1 = [h_{1,1}, \, \ldots, \, h_{1,N}]^{\Ttran} \in \mathbb{C}^N$ between the transmitter and RIS and of the channel $\vect{h}_2 = [h_{2,1}, \, \ldots, \, h_{2,N}]^{\Ttran} \in \mathbb{C}^N$ between the RIS and receiver.
To this end, we need to utilize the two-dimensional surface geometry.

\begin{figure}[t!]
	\centering 
	\begin{overpic}[width=.85\columnwidth,tics=10]{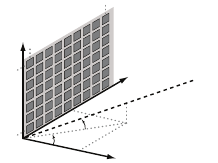}
		\put(54,1){\footnotesize $x$}
		\put(60,39){\footnotesize $y$}
		\put(9.5,57){\footnotesize $z$}
		\put(26.5,10){\footnotesize  $\varphi$}
		\put(40,17){\footnotesize  $\theta$}
		\put(5,11){\footnotesize  $1$}
		\put(1.5,41){\footnotesize  $N_{\mathrm{V}}$}
		\put(12.7,54){\footnotesize  $1$}
		\put(46,74){\footnotesize  $N_{\mathrm{H}}$}
		\put(80,31){\footnotesize  Multipath}
		\put(80,27){\footnotesize  component}
\end{overpic} 
	\caption{The 3D geometry of an RIS consisting of $N_{\mathrm{H}}$ elements per row and $N_{\mathrm{V}}$ elements per column.}\vspace{-0.3cm}
	\label{figure_geometric_setup}  
\end{figure}
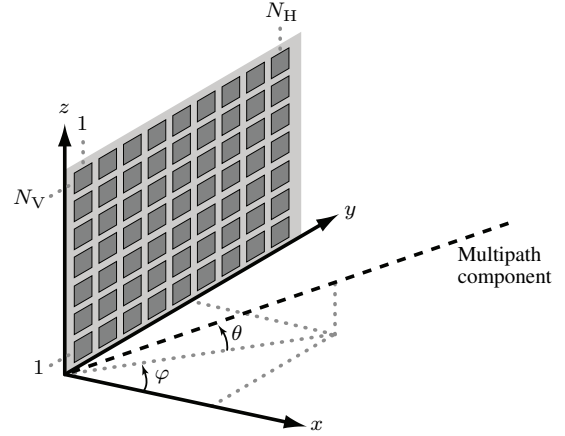

The RIS is a surface consisting of $N = N_{\mathrm{H}} N_{\mathrm{V}}$ elements which are deployed on a two-dimensional rectangular grid with $N_{\mathrm{H}}$ elements per row and $N_{\mathrm{V}}$ elements per column \cite{Renzo2020b}. 
The setup is illustrated in Fig.~\ref{figure_geometric_setup} in a three-dimensional (3D) space, where a local spherical coordinate system is defined with $\varphi$ being the azimuth angle and $\theta$ being the elevation angle.
Since the RIS is deployed in an isotropic scattering environment, the multipath components are uniformly distributed over the half-space in front of it, which is characterized by the probability density function (PDF)
\begin{equation} \label{eq:PDF-angles}
f(\varphi,\theta) = \frac{\cos(\theta)}{2\pi}, \quad \varphi \in \left[-\frac{\pi}{2},\frac{\pi}{2} \right], \theta \in \left[-\frac{\pi}{2},\frac{\pi}{2} \right].
\end{equation}

We assume each element has size $\dH \times \dV$, where $\dH$ is the horizontal width and $\dV$ is the vertical height. Hence, the area of an element is $A = \dH \dV$.
The elements are deployed edge-to-edge so the total area is $NA$. The elements
 are indexed row-by-row by $n\in[1,N]$, thus the location of the $n$th element with respect to the origin in Fig.~\ref{figure_geometric_setup} is
\begin{equation}
\vect{u}_n = [ 0, \, \,\, i(n) \dH,  \,\,\, j(n) \dV]^{\Ttran}
\end{equation}
where $i(n) =\mathrm{mod}(n-1,N_\mathrm{H})$ and $j(n) =\left\lfloor(n-1)/N_\mathrm{H}\right\rfloor$
are the horizontal and vertical indices of element $n$, respectively, on the two-dimensional grid. Notice that $\mathrm{mod}(\cdot,\cdot)$ denotes the modulus operation and $\lfloor \cdot \rfloor$ truncates the argument.

Suppose a plane wave with wavelength $\lambda$ is impinging on the RIS from the azimuth angle $\varphi$ and elevation angle $\theta$. The array response vector is then given by \cite[Sec.~7.3]{massivemimobook}
\begin{equation}\label{array-response}
\vect{a}(\varphi,\theta) = \left[e^{\imagunit\vect{k}(\varphi,\theta)^{\Ttran}\vect{u}_1},\dots,e^{\imagunit\vect{k}(\varphi,\theta)^{\Ttran}\vect{u}_N}\right]^{\Ttran}
\end{equation}
where $\vect{k}(\varphi, \theta)\in \mathbb{R}^{3 \times 1}$ is the wave vector
\begin{equation}
\vect{k}(\varphi, \theta) = \frac{2\pi}{\lambda}\left[\cos(\theta) \cos(\varphi), \,\,\, \cos(\theta) \sin(\varphi), \,\,\, \sin(\theta)\right]^{\Ttran}.
\end{equation}

\section{Rayleigh Fading Modeling}

In this section, we will derive the fading distribution for the channels $\vect{h}_1,\vect{h}_2$ and characterize their spatial channel correlation. The transmitter and receiver are assumed to be well separated so that their channels are independently distributed. We begin with analyzing $\vect{h}_1$. There are infinitely many multipath components in an isotropic scattering environment, but we begin by considering $L$ impinging plane waves:
\begin{equation}
\vect{h}_1 = \sum_{l=1}^L \frac{c_l}{\sqrt{L}} \vect{a}(\varphi_l,\theta_l)
\end{equation}
where $c_l/\sqrt{L} \in \mathbb{C}$ is the complex signal attenuation of the $l$th component, $\varphi_l$ is the azimuth angle-of-arrival, and $\theta_l$ is the elevation angle-of-arrival.
The attenuations $c_1,\ldots,c_L$ are i.i.d.~with zero mean and variance $A \mu_1$, where $A = \dH\dV$ is the area of an RIS element and $\mu_1$ is the average intensity attenuation.
The angles have the PDF $f(\varphi,\theta)$ in  \eqref{eq:PDF-angles}.

As $L \to \infty$, it follows from the central limit theorem that 
\begin{equation}
\vect{h}_1 \overset{d}{\rightarrow} \CN \left(  \vect{0},A \mu_1 \vect{R} \right)
\end{equation}
where the convergence is in distribution and the normalized spatial correlation matrix $\vect{R} \in \mathbb{C}^{N \times N}$ is computed as
\begin{equation}
\vect{R} = \frac{1}{A \mu_1} \mathbb{E} \left\{ \vect{h}_1 \vect{h}_1^{\Htran} \right\} =  \mathbb{E} \left\{ \vect{a}(\varphi,\theta) \vect{a}(\varphi,\theta)^{\Htran} \right\}.
\end{equation}
From~\eqref{array-response}, the $(n,m)$th element of $\vect{R}$ can be expanded as
\begin{align} \nonumber
&\!\!\!\![\vect{R}]_{n,m} =  \mathbb{E} \left\{ e^{\imagunit\vect{k}(\varphi,\theta)^{\Ttran}(\vect{u}_n-\vect{u}_m)} \right\}  \\
& \!\!\!\!= \mathbb{E} \left\{ e^{\imagunit \frac{2\pi}{\lambda} \left(  (i(n)-i(m)) \dH\!\cos(\theta) \sin(\varphi) + (j(n)-j(m)) \dV \!\sin(\theta) \right) } \right\}\!. \!\!\label{eq:R1_expression}
\end{align}

\begin{proposition} \label{prop:correlation-matrix}
With isotropic scattering in the half-space in front of the RIS, the spatial correlation matrix $\vect{R}$ has elements
\begin{equation} \label{eq:R1-expression}
[\vect{R}]_{n,m} =  \sinc \left( \frac{2 \| \vect{u}_n - \vect{u}_m \|}{\lambda} \right) \quad n,m= 1,\ldots,N
\end{equation}
where $\sinc(x) = \sin(\pi x)/ (\pi x)$ is the sinc function.
\end{proposition}
\begin{IEEEproof}
Consider two RIS elements $n$ and $m$ located on the same row, such that $i(n)=i(m)$ and $(j(n)-j(m)) \dV = \| \vect{u}_n - \vect{u}_m \| $. The expression in \eqref{eq:R1_expression} then simplifies to
\begin{align} \nonumber
[\vect{R}]_{n,m} & 
= \int_{-\pi/2}^{\pi/2} \int_{-\pi/2}^{\pi/2} e^{\imagunit \frac{2\pi}{\lambda}   \| \vect{u}_n - \vect{u}_m \| \sin(\theta) } f(\varphi,\theta)  d\theta d\varphi \\ \nonumber
&=   \int_{-\pi/2}^{\pi/2} e^{\imagunit \frac{2\pi}{\lambda}   \| \vect{u}_n - \vect{u}_m \| \sin(\theta) } \frac{\cos(\theta)}{2}   d\theta  \\
& =  \frac{\sin\left( \frac{2\pi}{\lambda}   \| \vect{u}_n - \vect{u}_m \|\right) }{ \frac{2\pi}{\lambda}   \| \vect{u}_n - \vect{u}_m \|} 
\end{align}
using Euler's formula. This expression is equal to \eqref{eq:R1-expression}. If the elements are not on the same row, we can rotate the coordinate system so that $\vect{u}_n - \vect{u}_m$ becomes a point on the new y-axis. By integrating over isotropic scatterers in the half-space in front of the RIS, we get the same result as above.
\end{IEEEproof}

Proposition~\ref{prop:correlation-matrix} characterizes the correlation matrix for the channel $\vect{h}_1$ from the transmitter to the RIS. As expected, it coincides with the Clarke's model for 3D spaces~\cite{Aulin1979}. Since the channel $\vect{h}_2$ from the RIS to the receiver is  subject to the same propagation conditions, it has the same distribution as $\vect{h}_1$, except for a different average intensity attenuation $\mu_2$.

\begin{corollary} \label{cor:Rayleigh-distributions}
In an isotropic scattering environment,  $\vect{h}_1,\vect{h}_2$ are independent and distributed as  
\begin{equation}
\vect{h}_i \sim \CN(\vect{0},A\mu_i\vect{R}) \quad i=1,2
\end{equation}
where the $(n,m)$th element of $\vect{R}$ is given by \eqref{eq:R1-expression}.
\end{corollary}

The average received signal power at the RIS is
\begin{equation} \label{eq:total-power}
\mathbb{E}\{ \| \vect{h}_1 s \|^2 \} = P A \mu_1 \tr \left( \vect{R} \right) = P \mu_1 \cdot \hspace{-0.4cm}\underbrace{NA}_{\textrm{Total RIS area}}
\end{equation}
since $\tr \left( \vect{R} \right) = N$ and it is proportional to the total RIS area $NA$. Hence, the propagation conditions are independent of the wavelength.
Since practical elements are sub-wavelength-sized $A \propto \lambda^2$ \cite{Ozdogan2019a,Tsilipakos2020a}, the number of elements $N$ needed to achieve a given total area $NA$ is inversely proportional to $\lambda^2$.

\subsection{Spatial Correlation}

The isotropic scattering environment gives rise to Rayleigh fading, as expected, but it will only be i.i.d.~Rayleigh fading if $\vect{R}$ is an identity matrix.
Proposition~\ref{prop:correlation-matrix} shows that the spatial correlation between two different RIS elements is a sinc-function of the physical distance between the elements divided by $\lambda/2$. Since the sinc-function is only zero for non-zero integer arguments, all the elements must be separated by $\lambda/2$ times different integers to achieve i.i.d.~fading. This is satisfied for any one-dimensional ULA with $\lambda/2$-spacing \cite{Marzetta2016a} or a two-dimensional triangular array with the right spacing between the three elements. None of these setups match with an RIS, which has correlation along all the diagonals and sub-$\lambda/2$ spacing.

\begin{corollary}\label{corr:spatial-correlation-RIS}
Any RIS deployed on a rectangular grid is subject to spatially correlated fading if $N_{\mathrm{H}}>1$ and $N_{\mathrm{V}}>1$.
\end{corollary}

This property holds for any practical RIS
since these are by definition two-dimensional. It also holds for other surface shapes than rectangles. The strength of the spatial correlation depends on the configuration. The eigenvalue spread of $\vect{R}$ is a common way to quantify the spatial correlation. In particular, one can consider its rank, i.e., $\rank(\vect{R})$. All the eigenvalues are equal in i.i.d.~Rayleigh fading and the rank is maximum, i.e., $\rank(\vect{R}) = N$. In correlated
channels, however, the rank can be smaller and the eigenvalues are non-identical.
\begin{figure}[t!]
	\centering 
	\begin{overpic}[width=1.05\columnwidth,tics=10]{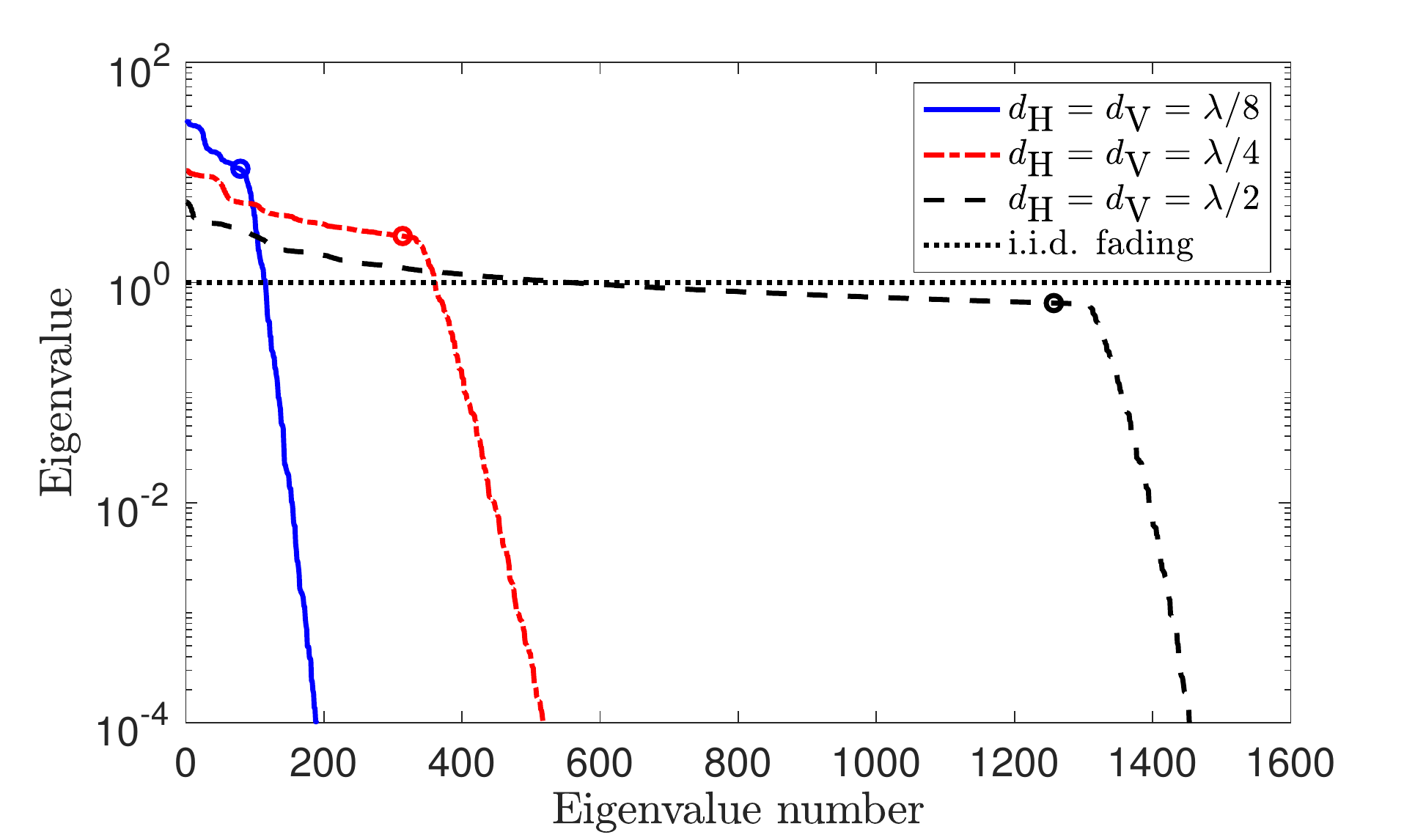}
\end{overpic} 
	\caption{The eigenvalues of $\vect{R}$ in decreasing order for an RIS with $N=1600$ and $\dH =\dV=d \in \{\lambda/8, \lambda/4, \lambda/2\}$.}\vspace{-0.3cm}
	\label{figure_simulationEigenvalues}  
\end{figure}
\begin{proposition} \label{proposition:rank}
As $N \to \infty$ and $A \to 0$ such that $NA \to \infty$, we have that 
\begin{equation}
\frac{\rank(\vect{R})}{\pi NA / \lambda^2} \to 1.
\end{equation}
\end{proposition}
\begin{IEEEproof}
Define the degrees of freedom (DoF) of the RIS as the number of non-zero eigenvalues of $\bf {R}$. Hence, we have that $\rm{DoF} = \rank(\vect{R})$. As $N \to \infty$ and $A \to 0$ such that $NA \to \infty$, the RIS becomes an infinitely large spatially-continuous electromagnetic aperture of rectangular geometry. Therefore, the proposition follows from \cite{Pizzo2020,Pizzo2020-SPAWC}, which prove that the $\rm{DoF}$ per m$^2$ in an isotropic environment is asymptotically (as the aperture size grows) equal to $\pi/ \lambda^2$.
\end{IEEEproof}

The practical interpretation of Proposition~\ref{proposition:rank} is that $\rank(\vect{R})$ can be approximated by $\pi NA / \lambda^2$ for a sufficiently large and dense RIS. This means that the $\pi NA / \lambda^2$ largest eigenvalues of $\vect{R}$ almost sum up to  $\tr(\vect{R})$, which is the sum of all the eigenvalues. This is illustrated in Fig.~\ref{figure_simulationEigenvalues}, which shows the eigenvalues of $\vect{R}$ in decreasing order in a setup with $N=1600$ elements ($N_{\mathrm{H}} = N_{\mathrm{V}} = 80$). We consider square elements of different size: $\dH =\dV = d \in \{\lambda/8, \lambda/4, \lambda/2\}$ with $A=d^2$.
The approximate rank $\pi N (d/\lambda)^2$ is indicated by circles on the curves. The figure shows that the first $\pi N (d/\lambda)^2$ eigenvalues are large but non-identical. After that, the eigenvalues quickly approach zero. The approximation is particularly good when $d$ is small, which is line with the proposition.
The i.i.d.~Rayleigh fading is also reported as reference. We can see that none of the considered cases resembles it. The case $\dH =\dV=\lambda/2$ is the closest one, but there are major differences: 25\% of the eigenvalues are larger than one, while 20\% of the eigenvalues are much smaller than one. Since an RIS is envisioned to be implemented with element sizes $d \in [\lambda/8, \lambda/4]$ \cite{Ozdogan2019a,Tsilipakos2020a}, we 
should expect the spatial correlation to be far from i.i.d.~fading.

\begin{remark}Proposition~\ref{proposition:rank} states that the eigenvectors associated with the $\pi NA / \lambda^2$ largest eigenvalues of $\bf R$ span the eigenspace where all the channel realizations reside.
This is a useful property during channel estimation. If $\vect{R}$ is known, the pilot signals can be transmitted along its eigenvectors and we can ignore those associated with the smallest eigenvalues \cite{Bjornson2010a}. Hence, it is sufficient to transmit approximately $\pi N A / \lambda^2$ pilot signals to estimate $\vect{h}_1$.
While the received power in \eqref{eq:total-power} only depends on the total area $NA$, the rank also depends on the wavelength $\lambda$. The rank increases with the carrier frequency and, thus, the pilot resources must also increase.\vspace{-0.3cm}
\end{remark}

\subsection{Comparison With the Kronecker model}

A so-called Kronecker model has been utilized to analyze the spatial correlation of planar arrays in previous works (e.g.,\cite{Ying2014a}). We will now compare it with the exact characterization provided by Proposition~\ref{prop:correlation-matrix}.
To this end, we enrich the notation by letting $\vect{R}_{(N_{\mathrm{H}},N_{\mathrm{V}})}$ denote the exact correlation matrix with  $N_{\mathrm{H}}$ elements per row and $N_{\mathrm{V}}$ columns.
The Kronecker model creates an approximate correlation matrix $\vect{R}_{(N_{\mathrm{H}},N_{\mathrm{V}})}^{\textrm{approx}} $ as \cite{Ying2014a}
\begin{equation} \label{eq:Kronecker-model}
\vect{R}_{(N_{\mathrm{H}},N_{\mathrm{V}})}^{\textrm{approx}} = \vect{R}_{(1,N_{\mathrm{V}})}  \kron \vect{R}_{(N_{\mathrm{H}},1)}
\end{equation}
where $\kron$ denotes the Kronecker product.
This is a combination of the spatial correlation matrix $\vect{R}_{(1,N_{\mathrm{V}})}\in \mathbb{C}^{N_{\mathrm{V}} \times N_{\mathrm{V}}}$ of a vertical ULA and $\vect{R}_{(N_{\mathrm{H}},1)}\in \mathbb{C}^{N_{\mathrm{H}} \times N_{\mathrm{H}}}$ of a horizontal ULA. 
The numerical results of \cite{Levin2010a,Ying2014a} show that the eigenvalue spectrum of $\vect{R}_{(N_{\mathrm{H}},N_{\mathrm{V}})}^{\textrm{approx}} $ matches quite well with $\vect{R}_{(N_{\mathrm{H}},N_{\mathrm{V}})}$ in a few simulation setups with small arrays.
However, the approximate equivalence breaks down immediately if an RIS with $\dH =\dV=\lambda/2$ is considered. In this case, both the vertical and horizontal ULAs have elements with $\lambda/2-$spacing and thus the spatial correlation between their respective elements is zero, i.e., $\vect{R}_{(1,N_{\mathrm{V}})} = {\bf I}_{N_{\mathrm{V}}}$ and $\vect{R}_{(1,N_{\mathrm{H}})} = {\bf I}_{N_{\mathrm{H}}}$. We have that $\vect{R}_{(N_{\mathrm{H}},N_{\mathrm{V}})}^{\textrm{approx}} = {\bf I}_N$, thus the approximation in~\eqref{eq:Kronecker-model} gives rise to i.i.d.~Rayleigh fading if $\dH =\dV=\lambda/2$. This is in contrast to Corollary~\ref{corr:spatial-correlation-RIS} and makes the Kronecker model inappropriate in these conditions. Notice also that $\rank(\vect{R}_{(1,N_{\mathrm{V}})} ) = \rank( \vect{R}_{(N_{\mathrm{H}},1)}) \approx 2 NA/\lambda$ as the ULAs grow large and inter-element spacing becomes smaller.\footnote{This can be proved following the same steps of the proof in Proposition~\ref{proposition:rank} using the results in \cite[Sec. III.A]{Pizzo2020-SPAWC}.} Under the conditions of Proposition~\ref{proposition:rank}, we then have that  
\begin{equation}
\rank(\vect{R}_{(N_{\mathrm{H}},N_{\mathrm{V}})}^{\textrm{approx}}) = \rank(\vect{R}_{(1,N_{\mathrm{V}})} ) \rank( \vect{R}_{(N_{\mathrm{H}},1)}) \approx \frac{4 NA}{\lambda^2}
\end{equation}
while $\rank(\vect{R}_{(N_{\mathrm{H}},N_{\mathrm{V}})}) \approx \pi NA/\lambda^2$; that is, the rank is miscalculated by a factor $\pi/4 < 1$ \cite{Pizzo2020-SPAWC}. Hence, the Kronecker model does not capture the basic properties of a large RIS.

\begin{figure}[t!]
	\centering 
	\begin{overpic}[width=1.05\columnwidth,tics=10]{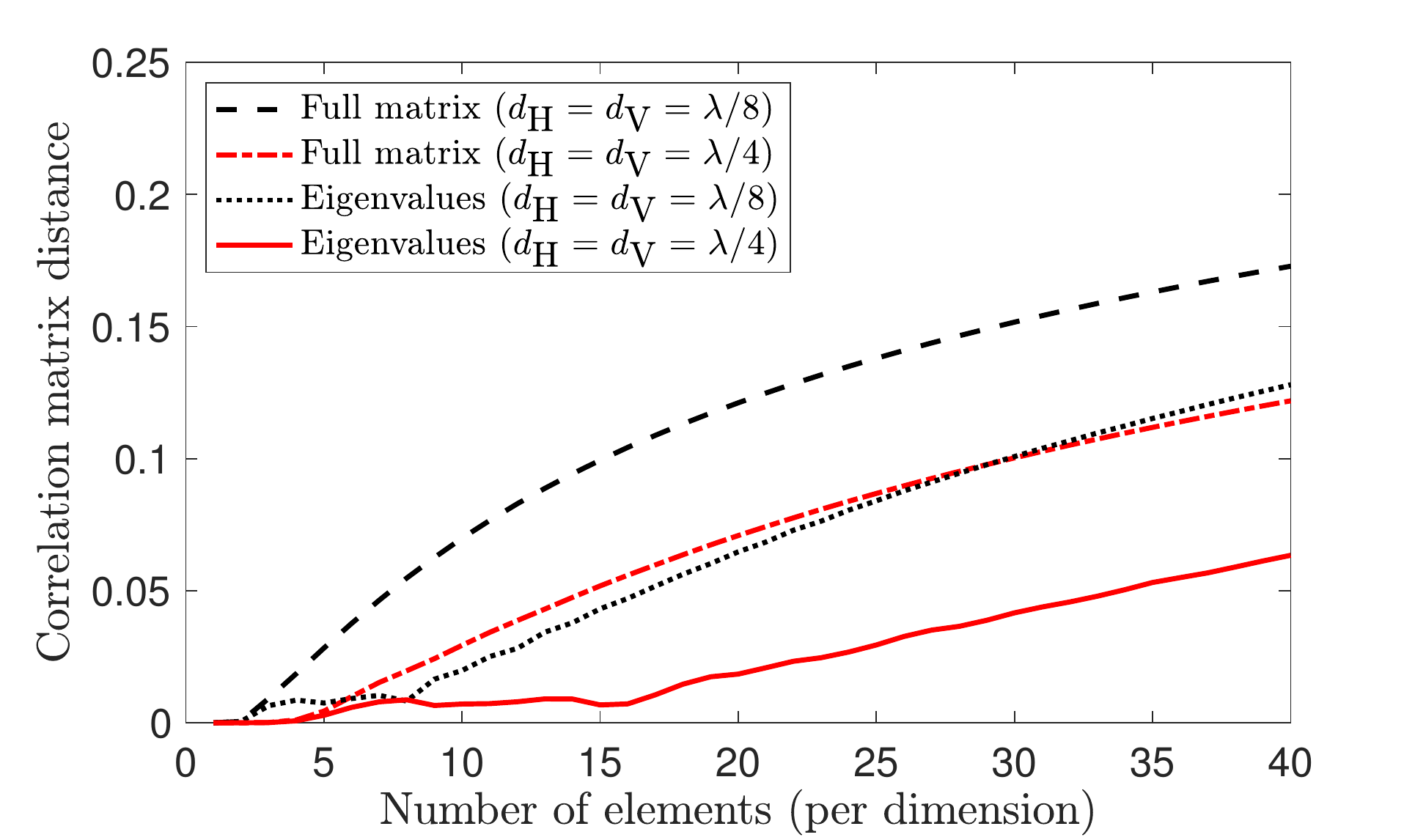}
\end{overpic} 
	\caption{The distance between the exact correlation matrix and Kronecker model in \eqref{eq:Kronecker-model} for varying $N_{\mathrm{H}}=N_{\mathrm{V}} \in [1,40]$.}\vspace{-0.2cm}
	\label{simulationEigenvalues_approx}  
\end{figure}

Moreover, the eigenvectors are not matching, thus $\vect{R}_{(N_{\mathrm{H}},N_{\mathrm{V}})}$ and  $\vect{R}_{(N_{\mathrm{H}},N_{\mathrm{V}})}^{\textrm{approx}} $ are more different than their eigenvalue spectra reveal. This is illustrated in Fig.~\ref{simulationEigenvalues_approx} that compares the exact and approximate matrices using the correlation matrix distance \cite{Herdin2005a}, which is a value between 0 and 1.\footnote{For two correlation matrices $\vect{A},\vect{B}$ of matching size, the correlation matrix distance in \cite{Herdin2005a} is $1-\tr(\vect{A}\vect{B})/ (\| \vect{A} \| \| \vect{B} \|)$ using the Frobenius norm.} There are curves for  $\dH =\dV \in \{\lambda/8, \lambda/4\}$ considering either the full matrices or only diagonal matrices containing the ordered eigenvalues. Fig.~\ref{simulationEigenvalues_approx}  shows that the distance (i.e., approximation error) increases with the RIS size, and that the full matrices are much more different than their eigenvalue spectra.

\section{Channel Hardening}

Channel fading has a negative impact on the communication performance due to the signal-to-noise ratio (SNR) variations that it creates. MIMO channels generally provide spatial diversity that can reduce such variations. In particular, i.i.d.~Rayleigh fading channels give rise to so-called \emph{channel hardening}, where the SNR variations average out (in relative terms) as the number of antennas increases \cite{Hochwald2004a,massivemimobook}. We will now provide a new general definition of channel hardening that can be utilized in RIS-aided communications.

With the optimal phase-configuration $\phi_n = \arg(h_{1n}h_{2n})-\arg(h_{\mathrm{d}})$
\cite{Huang2018a,Ozdogan2019a}, the \emph{instantaneous} SNR of the system in \eqref{eq:received-signal1} is \vspace{-2mm}
\begin{equation}
\mathrm{SNR}_{\vect{h}_1,\vect{h}_2,h_{\mathrm{d}}} = \frac{P}{\sigma^2} \left(\sum_{n=1}^N |h_{1n} h_{2n}|  + | h_{\mathrm{d}}| \right)^2.
\end{equation}
This SNR plays a key role in fast fading scenarios, where the ergodic rate is $\mathbb{E} \left\{ \log_2 \left( 1 +\mathrm{SNR}_{\vect{h}_1,\vect{h}_2,h_{\mathrm{d}}}  \right) \right\}$.
It is important also in slow fading scenarios, where the outage probability for a rate $R$ is $\Pr \left\{ \log_2 \left( 1 +\mathrm{SNR}_{\vect{h}_1,\vect{h}_2,h_{\mathrm{d}}}  \right) < R \right\}$.
The randomness of $\vect{h}_1,\vect{h}_2,h_{\mathrm{d}}$  determines the performance in both cases.

\begin{definition} \label{def:hardening}
Asymptotic channel hardening occurs in an RIS-aided communication system if
\begin{equation}
\frac{\mathrm{SNR}_{\vect{h}_1,\vect{h}_2,h_{\mathrm{d}}} }{N^2} \to \textrm{constant} \quad \textrm{as } N \to \infty.
\end{equation}
\end{definition}

This definition involves convergence of sequences of the random variables $\{h_{\mathrm{d}}, h_{1n},h_{2n}: n=1,\ldots,N\}$, which can be defined differently (e.g., convergence in probability or almost surely \cite{massivemimobook}). The type of convergence is irrelevant in this context since it is the behavior for large but finite $N$ that matters; the channel models break down if we physically let $N \to \infty$ \cite{Bjornson2020b}.
The practical interpretation of channel hardening is that the random $\mathrm{SNR}_{\vect{h}_1,\vect{h}_2,h_{\mathrm{d}}}$ is approximately equal to $N^2$ times a deterministic constant when $N$ is large. The quadratic scaling makes the behavior very different from Massive MIMO and is called the ``square law'' \cite{Wu2019a}. We will prove that channel hardening appears with the new fading model.

\begin{lemma}\label{lemma-LLN-correlated}
Let $\{ X_n \}$ be a sequence of random variables with mean value $A$, bounded variances, and covariance $\mathrm{Cov}\{X_i,X_j\} \to 0$ when $|i-j| \to \infty$, then
\begin{equation}
\frac{1}{N} \sum_{n=1}^{N} X_n \to A
\end{equation}
with convergence in probability.
\end{lemma}
\begin{IEEEproof}
This is a special case of \cite[Ex.~254]{CacoullosT1989EiP}.
\end{IEEEproof}

\begin{figure} 
        \centering\vspace{-2mm}
                \begin{subfigure}[t]{\columnwidth} \centering 
	\begin{overpic}[width=1.07\columnwidth,tics=10]{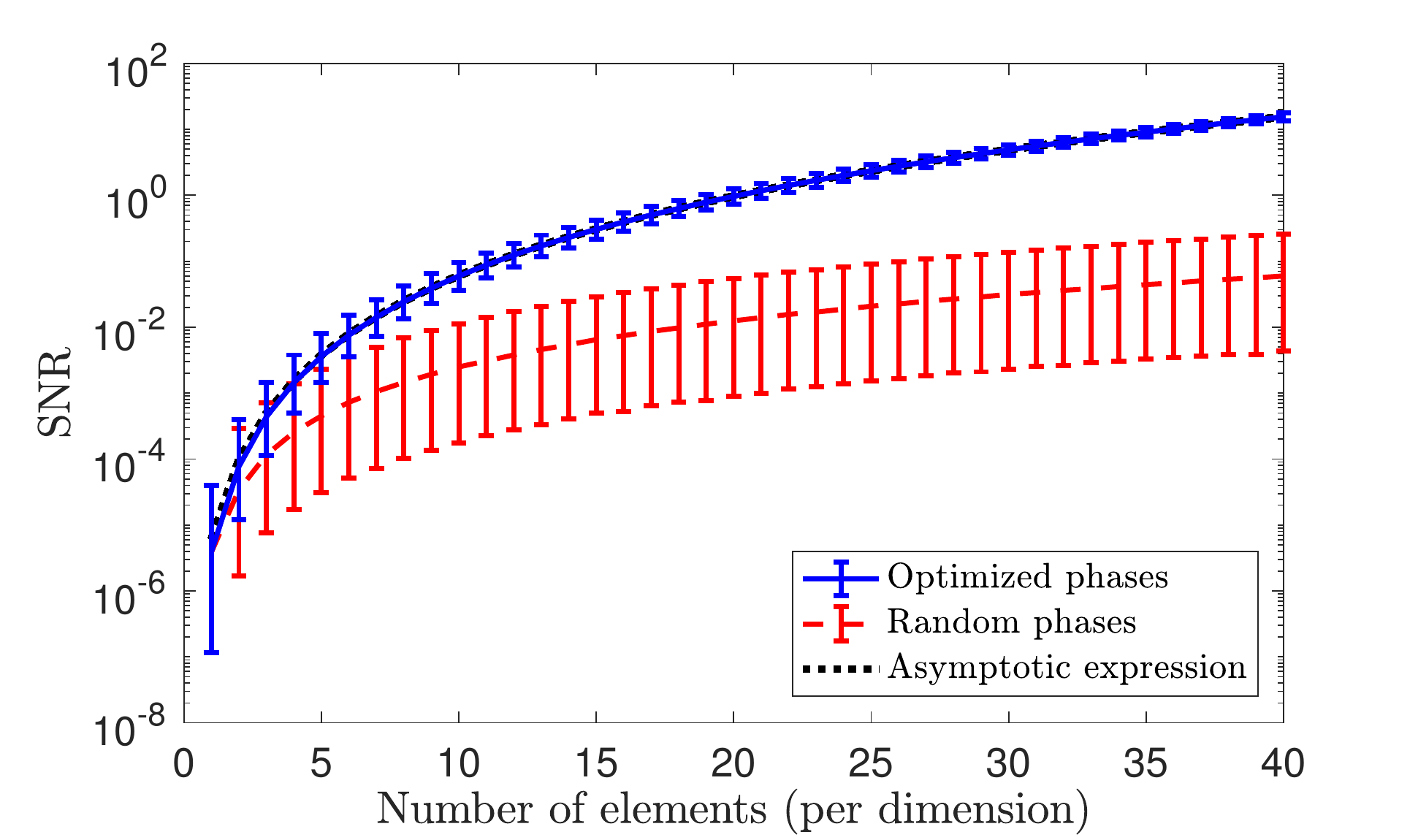}
\end{overpic} 
                \caption{Without direct path: $\beta_{\mathrm{d}} = -\infty$\,dB} \vspace{+1mm}
                \label{simulationHardening-NoDirectPath} 
        \end{subfigure} 
        \begin{subfigure}[t]{\columnwidth} \centering 
	\begin{overpic}[width=1.07\columnwidth,tics=10]{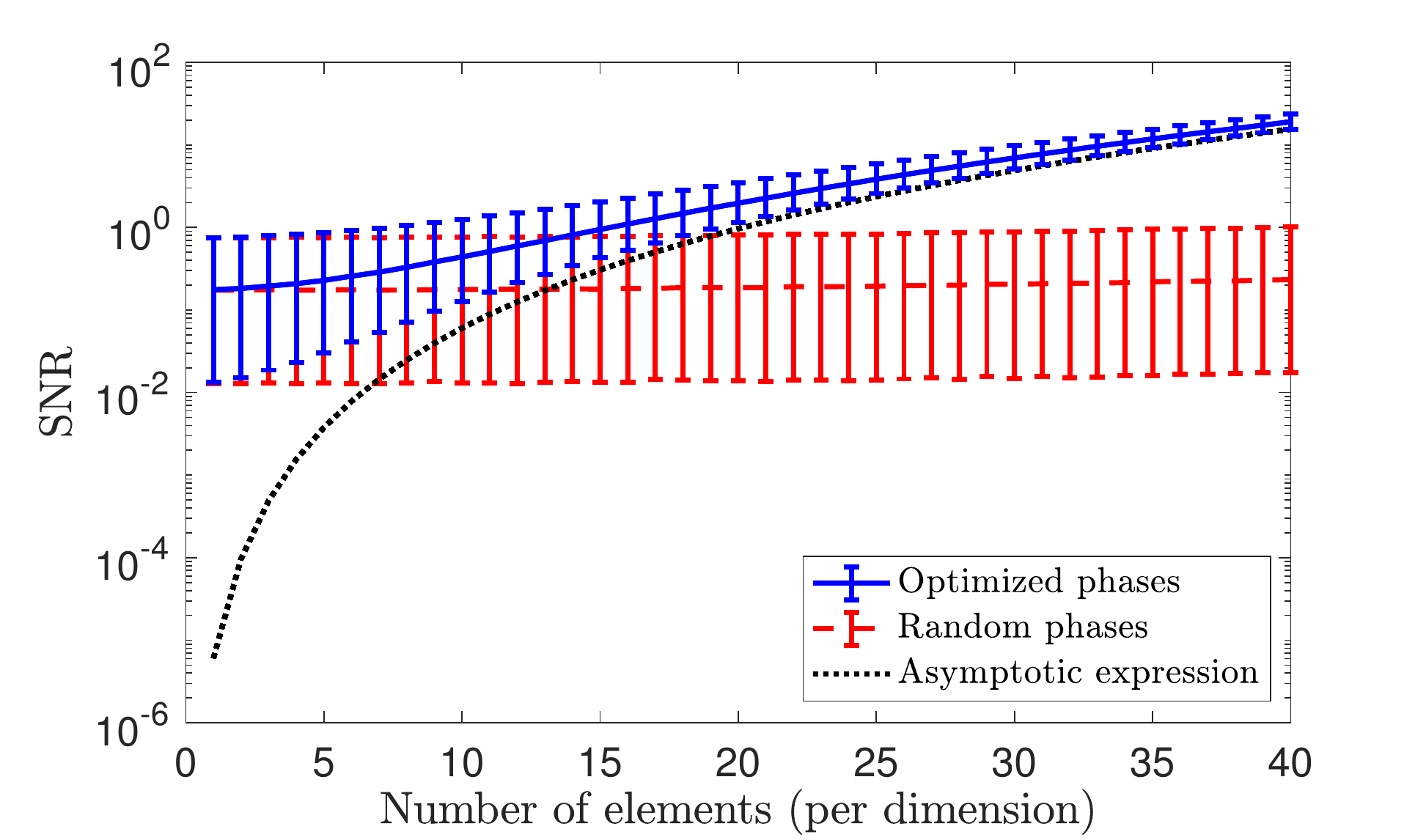}
\end{overpic} 
                \caption{With direct path: $\beta_{\mathrm{d}} = -130$\,dB} 
                \label{simulationHardening-WithDirectPath} 
        \end{subfigure} 
        \caption{The SNR achieved with an optimized RIS and with random phase-shifts for varying $N_{\mathrm{H}}=N_{\mathrm{V}} \in [1,40]$. Convergence to the asymptotic expression in \eqref{eq:SNR-approx} is illustrated with or without the direct path.}\vspace{-0.3cm}
	\label{simulationHardening} 
\end{figure}

\begin{proposition} \label{prop:hardening}
In an isotropic scattering environment with $\vect{h}_1$ and $\vect{h}_2$ being independent and distributed as in Corollary~\ref{cor:Rayleigh-distributions}, it holds that
\begin{equation} \label{eq:hardening}
\frac{\left(\sum_{n=1}^N |h_{1n} h_{2n}|  + | h_{\mathrm{d}}| \right)^2}{N^2 } \to A^2 \mu_1 \mu_2 \frac{\pi^2}{16} \quad \textrm{as } N \to \infty
\end{equation}
where the convergence is in probability.
\end{proposition}
\begin{IEEEproof}
Corollary~\ref{cor:Rayleigh-distributions} implies $|h_{1n}| \sim \mathrm{Rayleigh}(\sqrt{A\mu_1/2})$ and $|h_{2n}| \sim \mathrm{Rayleigh}(\sqrt{A\mu_2/2})$. Due to their mutual independence, it follows that $\mathbb{E}\{ |h_{1n} h_{2n}| \} = A \pi \sqrt{\mu_1 \mu_2}/4$ and that the variance is bounded.
A consequence of Proposition~\ref{prop:correlation-matrix} is that the covariance between $h_{1n}$ and $h_{1m}$ goes to zero as $|n-m| \to \infty$, thus we can invoke Lemma~\ref{lemma-LLN-correlated} to obtain
\begin{equation}
\frac{1}{N} \sum_{n=1}^N |h_{1n} h_{2n}| \to \frac{A \pi \sqrt{\mu_1 \mu_2}}{4}
\end{equation}
with convergence in probability. It then follows that \vspace{-2mm}
\begin{align} \nonumber
&\frac{1}{N^2} \left(\sum_{n=1}^N |h_{1n} h_{2n}|  + | h_{\mathrm{d}}| \right)^2 \\
&= \left( \frac{1}{N} \sum_{n=1}^N |h_{1n} h_{2n}|  + \frac{1}{N} | h_{\mathrm{d}}| \right)^2 \to \left(\frac{ A \pi \sqrt{\mu_1 \mu_2}}{4} \right)^2 \label{eq:proof-hardening}
\end{align}
by exploiting that $ | h_{\mathrm{d}}| /N \to 0$. This is equivalent to   \eqref{eq:hardening}.
\end{IEEEproof}

This proposition shows that the random SNR in an isotropic scattering environment can be approximated by a deterministic term as
\begin{equation} \label{eq:SNR-approx}
\mathrm{SNR}_{\vect{h}_1,\vect{h}_2,h_{\mathrm{d}}} \approx  \frac{P}{\sigma^2}  \mu_1 \mu_2 \left(\frac{\pi}{4} AN\right)^2
\end{equation}
when the RIS is sufficiently large. This deterministic approximation coincides with that in \cite[Prop. 2]{Zhang2019x}, which is obtained under the (unrealistic) assumption of i.i.d. Rayleigh fading and no direct path. 
This is natural since the average received power is equal, but the convergence is very different: the spatial correlation between $h_{1n}$ and $h_{1m}$ in the Rayleigh fading model in Corollary~\ref{cor:Rayleigh-distributions} goes to zero as $|n-m|$ increases and the direct path disappears since it is independent of $N$. 
Although \eqref{eq:SNR-approx} does not depend on the strength of the direct path, $\beta_d$, that component determines how many elements are needed before the approximation can be applied. This is because the path via the RIS must be much stronger than it.

To exemplify this property, we consider a setup where $A \mu_1 = A \mu_2 = -75$\,dB,  $\dH =\dV = \lambda/4$, and $P/\sigma^2 = 124$\,dB (which corresponds to transmitting 1\,W over 10 MHz of bandwidth, with 10 dB noise figure). We assume a square RIS with $N_{\mathrm{H}} = N_{\mathrm{V}}$ elements and Fig.~\ref{simulationHardening} shows the SNR as a function of $N_{\mathrm{H}}$ without and with the direct path. In the latter case, we assume $\beta_{\mathrm{d}} = -130$\,dB. There is one curve for an RIS with optimal phases and one with random phases \cite{Zhang2019x}.\footnote{Notice that choosing $\boldsymbol{\Phi} = {\bf I}e^{-\imagunit \phi}$ with $\phi$ being an arbitrary phase provides the same results as the case with random phases $\{\phi_1,\ldots,\phi_N\}$.} The curves show the median value and the bars indicate the interval where 90\% of the random realizations appears (computed based on 50000 Monte Carlo trials). With optimized phases, the SNR grows as $N^2=N_{\mathrm{H}}^4$. This is particularly evident in Fig.~\ref{simulationHardening-NoDirectPath} where the direct path is absent while it is observed in Fig.~\ref{simulationHardening-WithDirectPath} only when the RIS path is stronger than the direct path. This happens for $N_{\mathrm{H}}\geq 
(\beta_d / ( \mu_1 \mu_2 A^2))^{1/4} \approx 12$. The dotted curve shows the deterministic approximation in \eqref{eq:SNR-approx}.
The instantaneous SNR matches well with the dotted curve for $N_{\mathrm{H}}\geq 10$ in Fig.~\ref{simulationHardening-NoDirectPath}, in the sense that the median becomes closer and the random variations reduce (in relative terms). A larger number $N_{\mathrm{H}}\geq 25$ is needed in Fig.~\ref{simulationHardening-WithDirectPath} because of the presence of the direct path. These are examples of the channel hardening proved by Proposition 3.
In contrast, the random SNR variations in the case with random phases remain large in both cases 
since there is no hardening. When the direct path is present, the SNR increases very slowly with $N$.
Hence, an RIS must be properly configured to benefit from the channel hardening and SNR gain.
\vspace{-2mm}

\section{Conclusions}

The channel fading in RIS-aided communications will always be spatially correlated, thus we discourage from using the i.i.d.~Rayleigh fading model. The asymptotic SNR limit is equal, but the convergence rate and rank of the spatial correlation matrices are different. We have provided an accurate channel model for isotropic scattering and characterized its properties, including rank and channel hardening.
The derived channel properties also apply for Massive MIMO arrays with the same form factor, called holographic MIMO \cite{Pizzo2020,Huang2020a}.

\vspace{-0.2cm}
\bibliographystyle{IEEEtran}

\bibliography{IEEEabrv,refs}

\end{document}